\documentstyle{mn}

\input epsf

\newif\ifAMStwofonts

\def\fun#1#2{\lower3.6pt\vbox{\baselineskip0pt\lineskip.9pt
        \ialign{$\mathsurround=0pt#1\hfill##\hfil$\crcr#2\crcr\sim\crcr}}}

\renewcommand\({\left(}
\renewcommand\){\right)}

\newcommand\eq[1]{Eq.~(\ref{#1})}
\newcommand\eqs[2]{Eqs.~(\ref{#1}) and (\ref{#2})}

\newcommand\eqsss[4]{Eqs.~(\ref{#1}), (\ref{#2}), (\ref{#3})
and (\ref{#4})}

\newcommand\ee{\end{equation}}
\newcommand\be{\begin{equation}}
\newcommand\eea{\end{eqnarray}}
\newcommand\bea{\begin{eqnarray}}

\newcommand\muK{\,\mu\mbox{K}}

\newcommand\GeV{\,\mbox{GeV}}

\newcommand\Mpc{\,\mbox{Mpc}}

\newcommand\mpl{M_{\rm P}}

\newcommand\lsim{\mathrel{\rlap{\lower4pt\hbox{\hskip1pt$\sim$}}
    \raise1pt\hbox{$<$}}}
\newcommand\gsim{\mathrel{\rlap{\lower4pt\hbox{\hskip1pt$\sim$}}
    \raise1pt\hbox{$>$}}}

\newcommand\diff{\mbox d}

\def\dslash{\not{\hbox{\kern-2pt $\partial$}}}
\def\Dslash{\not{\hbox{\kern-4pt $D$}}}
\def\Oslash{\not{\hbox{\kern-4pt $O$}}}
\def\Qslash{\not{\hbox{\kern-4pt $Q$}}}
\def\pslash{\not{\hbox{\kern-2.3pt $p$}}}
\def\kslash{\not{\hbox{\kern-2.3pt $k$}}}
\def\qslash{\not{\hbox{\kern-2.3pt $q$}}}

 \newtoks\slashfraction
 \slashfraction={.13}
 \def\slash#1{\setbox0\hbox{$ #1 $}
 \setbox0\hbox to \the\slashfraction\wd0{\hss \box0}/\box0 }
 
\def\ee{\end{equation}}
\def\be{\begin{equation}}

\def\calp{{\cal P}}
\def\calr{{\cal R}}
\def\calpr{{\calp_\calr}}

\newcommand\sub[1]{_{\rm #1}}
\newcommand\su[1]{^{\rm #1}}

\newcommand\omb{\Omega\sub b}

\ifoldfss
  \ifCUPmtlplainloaded \else
    \NewTextAlphabet{textbfit} {cmbxti10} {}
    \NewTextAlphabet{textbfss} {cmssbx10} {}
    \NewMathAlphabet{mathbfit} {cmbxti10} {} 
    \NewMathAlphabet{mathbfss} {cmssbx10} {} 
  \fi
  \ifAMStwofonts
    \ifCUPmtlplainloaded \else
      \NewSymbolFont{upmath} {eurm10}
      \NewSymbolFont{AMSa} {msam10}
      \NewMathSymbol{\upi}     {0}{upmath}{19}
      \NewMathSymbol{\umu}     {0}{upmath}{16}
      \NewMathSymbol{\upartial}{0}{upmath}{40}
      \NewMathSymbol{\leqslant}{3}{AMSa}{36}
      \NewMathSymbol{\geqslant}{3}{AMSa}{3E}

      \let\geq=\geqslant 
    \fi
  \fi
\fi 

\ifnfssone
  \newmathalphabet{\mathit}
  \addtoversion{normal}{\mathit}{cmr}{m}{it}
  \addtoversion{bold}{\mathit}{cmr}{bx}{it}
  \newmathalphabet{\mathbfit} 
  \addtoversion{normal}{\mathbfit}{cmr}{bx}{it}
  \addtoversion{bold}{\mathbfit}{cmr}{bx}{it}
  \newmathalphabet{\mathbfss} 
  \addtoversion{normal}{\mathbfss}{cmss}{bx}{n}
  \addtoversion{bold}{\mathbfss}{cmss}{bx}{n}
  \ifAMStwofonts
    \ifCUPmtlplainloaded \else
      \UseAMStwoboldmath
      \makeatletter
      \new@mathgroup\upmath@group
      \define@mathgroup\mv@normal\upmath@group{eur}{m}{n}
      \define@mathgroup\mv@bold\upmath@group{eur}{b}{n}
      \edef\UPM{\hexnumber\upmath@group}
      \new@mathgroup\amsa@group
      \define@mathgroup\mv@normal\amsa@group{msa}{m}{n}
      \define@mathgroup\mv@bold\amsa@group{msa}{m}{n}
      \edef\AMSa{\hexnumber\amsa@group}
      \makeatother
      \mathchardef\upi="0\UPM19
      \mathchardef\umu="0\UPM16
      \mathchardef\upartial="0\UPM40
      \mathchardef\leqslant="3\AMSa36
      \mathchardef\geqslant="3\AMSa3E

      \let\geq=\geqslant 
    \fi
  \fi
\fi 

\ifnfsstwo
  \DeclareMathAlphabet{\mathbfit}{OT1}{cmr}{bx}{it}
  \SetMathAlphabet\mathbfit{bold}{OT1}{cmr}{bx}{it}
  \DeclareMathAlphabet{\mathbfss}{OT1}{cmss}{bx}{n}
  \SetMathAlphabet\mathbfss{bold}{OT1}{cmss}{bx}{n}
  \ifAMStwofonts
    \ifCUPmtlplainloaded \else
      \DeclareSymbolFont{UPM}{U}{eur}{m}{n}
      \SetSymbolFont{UPM}{bold}{U}{eur}{b}{n}
      \DeclareSymbolFont{AMSa}{U}{msa}{m}{n}
      \DeclareMathSymbol{\upi}{0}{UPM}{"19}
      \DeclareMathSymbol{\umu}{0}{UPM}{"16}
      \DeclareMathSymbol{\upartial}{0}{UPM}{"40}
      \DeclareMathSymbol{\leqslant}{3}{AMSa}{"36}
      \DeclareMathSymbol{\geqslant}{3}{AMSa}{"3E}

      \let\geq=\geqslant 
    \fi
  \fi
\fi 

\ifCUPmtlplainloaded \else
  \ifAMStwofonts \else 
    \def\upi{\pi}
    \def\umu{\mu}
    \def\upartial{\partial}
  \fi
\fi

\begin{document}

\title{Global fits for the spectral index of the cosmological
curvature perturbation }
\author[L. Covi and D. H. Lyth]{Laura Covi$^{1}$ and David H. Lyth$^2$,  \\
$^1$DESY Theory Group, Notkestrasse 85, D-22603 Hamburg, Germany\\
$^2$Physics Department, Lancaster University Lancaster LA1 4YB, 
Great Britain}


\maketitle

\label{firstpage}

\begin{abstract}
Best-fit values of the spectral index of the curvature perturbation are 
presented, assuming the $\Lambda$CDM cosmology. Apart from the spectral 
index, the parameters are the Hubble parameter, the total  matter density
and the baryon density. The data points are intended to represent all 
measurements which are likely to significantly affect the result. The cosmic 
microwave anisotropy is represented by the COBE normalization, and heights 
of the first and second  peaks given by the latest Boomerang and Maxima data.
The slope of the galaxy correlation function and the matter density contrast
on the $8h^{-1}\Mpc$ scale are each represented by a data point,
as are the 
  expected values of the Hubble parameter  and 
matter  density.
 The `low-deuterium' nucleosynthesis value 
of the baryon density provides a final data point,
 the fit giving  a value about one 
standard deviation higher.  The reionization 
epoch  is calculated from the model by  assuming that it 
corresponds to the collapse of a  fraction $f\gsim 10^{-4}$ of matter. We 
consider the case of a scale-independent spectral index, and also the 
scale-dependent spectral index predicted by running mass models of inflation.
In the former case, the result is compared with the prediction of models of 
inflation based on effective field theory, in which the field value is small 
on the Planck scale. Detailed comparison is made with other fits,
and other approaches to the comparison with theory.
\end{abstract}


\section{Introduction}
The spectral index $n$, giving the scale dependence of 
 the spectrum $\calp_\calr$ of the  primordial curvature perturbation,
 will be 
 a powerful discriminator between models of inflation when it is 
accurately determined. Just before the release of the latest
Boomerang \cite{boom} and
Maxima data \cite{maxima} on the cosmic microwave background 
(cmb) anisotropy, we
reported \cite{cl99}
a global fit to the key pieces of available data. We considered
the case of a practically scale-independent spectral index, comparing
the best-fit value with some  models of inflation based on 
effective field theory.
We went on to consider the running mass models,
corresponding to a spectral index with strong  scale dependence,
and demonstrated  that such scale dependence was allowed by the data.

In the present paper, we update the fit by including the Boomerang and Maxima
results for  the height of the second peak of the cmb anisotropy,
for consistency taking the height of the first peak from the same source.
The best-fit values of the spectral index and other parameters are 
different from the previous case, but not dramatically so, while 
the value of   $\chi^2$, though higher,  is still acceptable.
We consider  in some   detail the implication of our
results for  some  models of inflation based on effective field theory,
 drawing a distinction between
such models and  ad hoc parameterizations of the potential. 
In the running mass model, strong scale dependence of the spectral index is
still permitted by the data.

As with our previous fit, we assume 
  the $\Lambda$CDM cosmology,
in which the Universe is flat and the dark matter is cold. 
Flatness is the naive  prediction of inflation, and there is at
present no firm motivation for considering modifications of the simplest
 dark matter hypothesis. The model, then, consists of the $\Lambda$CDM 
cosmology, the assumption that a gaussian 
 primordial curvature perturbation is the 
only one, and the assumption about reionization that we shall discuss shortly.

\section{The fit}

\begin{center}
\begin{figure*}
\leavevmode
\epsfysize=7.5cm\epsfbox{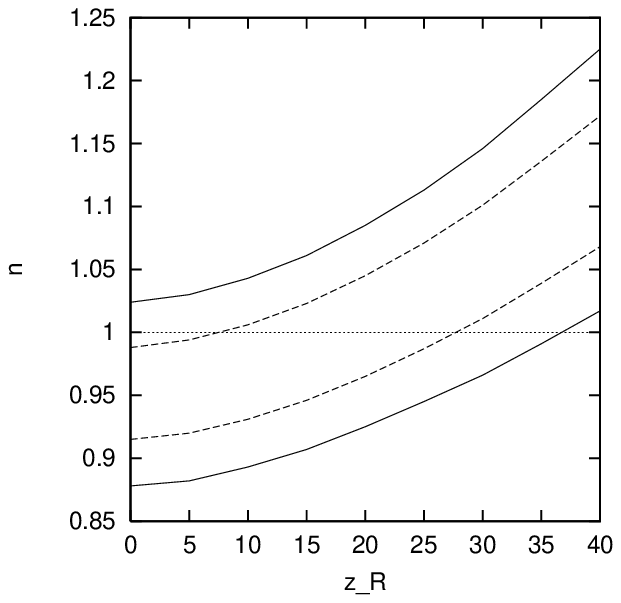}
\epsfysize=7.5cm\epsfbox{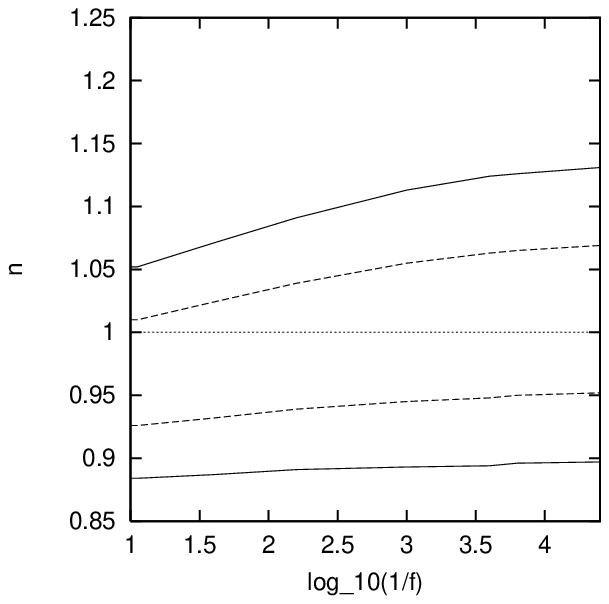}\\
\leavevmode
\epsfysize=7.5cm\epsfbox{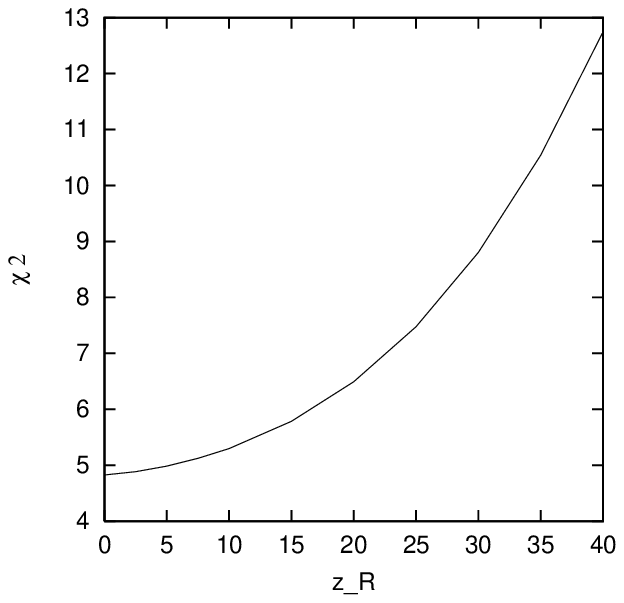}
\epsfysize=7.5cm\epsfbox{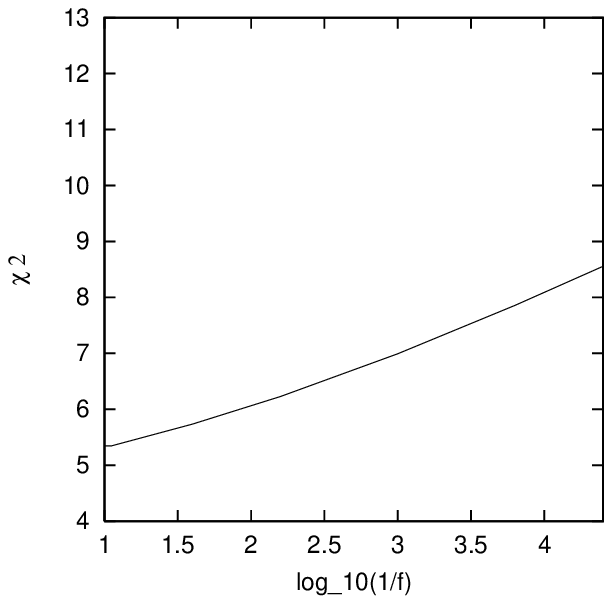}\\
\caption{The top panels show 
 nominal 1- and 2-$\sigma$ bounds on $n$. In the left-hand
panel the reionization epoch $z\sub R$ is fixed. In the right-hand panel,
is fixed instead the fraction  $f$ of matter which is assumed to have collapsed when at the epoch of reionization.
(The corresponding reionization redshift, at best fit,
 is  in the range $10$
to $26$.) The bottom panels show $\chi^2$, with
three degrees of freedom.}
\label{fig1}
\end{figure*}
\end{center}

\begin{center}
\begin{figure*}
\leavevmode
\epsfysize=7.5cm \epsfbox{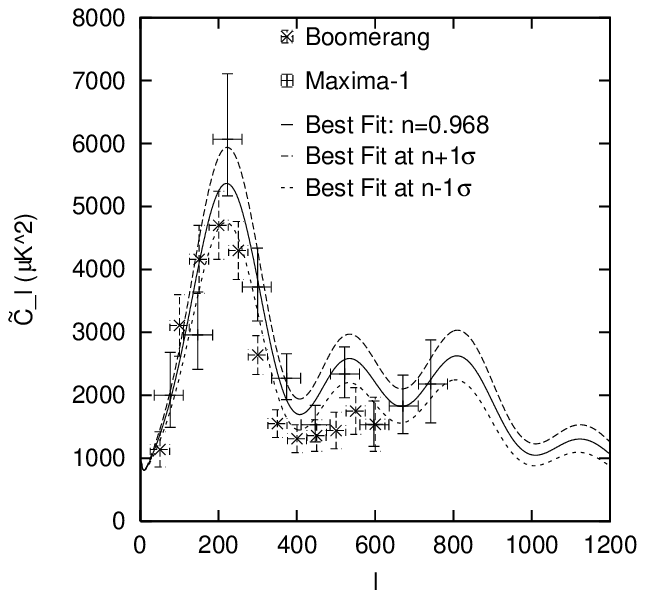}
\epsfysize=7.5cm \epsfbox{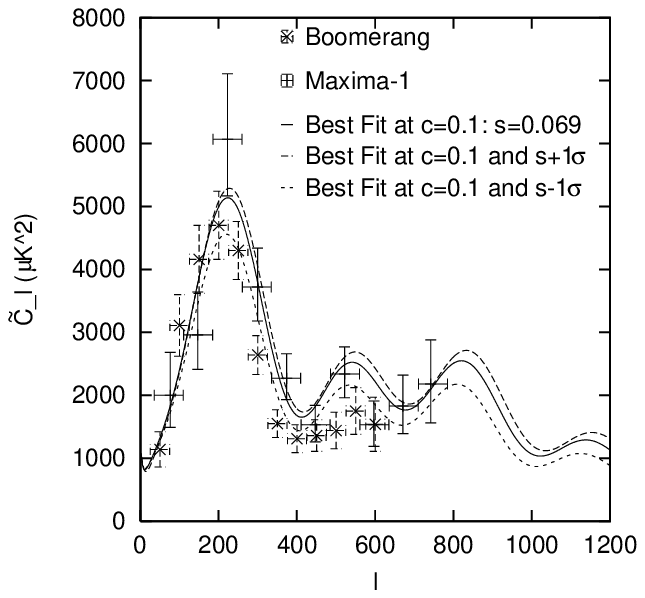}\\
\leavevmode
\epsfysize=7.5cm \epsfbox{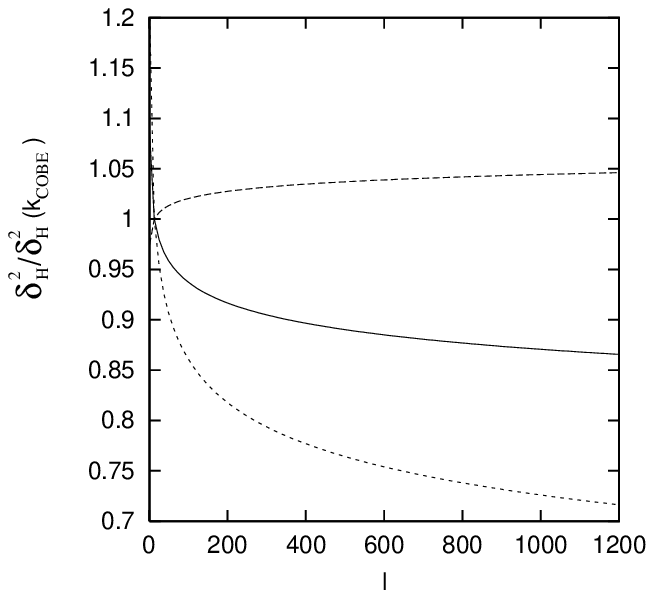}
\epsfysize=7.5cm \epsfbox{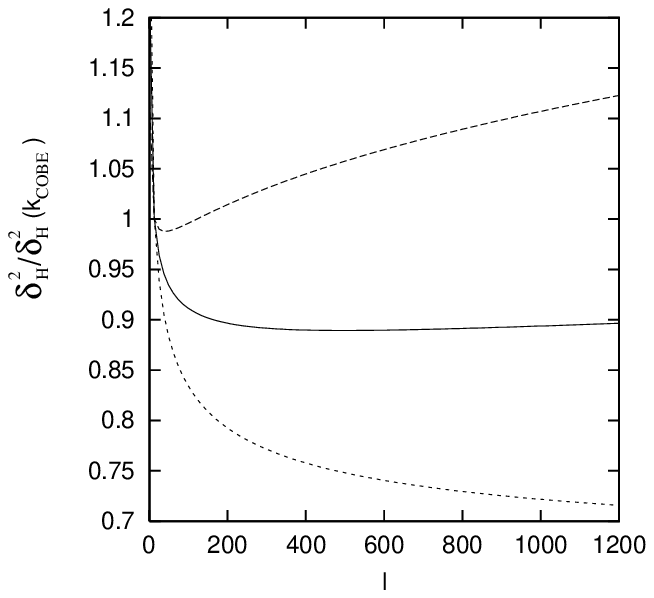}\\  
\caption{
Curves correspond to best fit $\pm$ 1-$\sigma$ for $f\simeq 1$.
The top panels show the cmb anisotropy $\widetilde C_\ell$, for 
the case of scale-independent spectral index (left panel) and for the 
 running mass model with coupling $c=0.1$ (right panel). 
The error bars do not include systematic errors; 
taking these as the calibration uncertainties they are $20\%$ for Boomerang
and $8\%$ for Maxima. The fit used only the two
data points nearest each of the peaks (one each from Boomerang and Maxima)
and added in quadrature the systematic errors. 
Other data sets around the
first peak (not shown) span a wider range and their rejection in favour
of Boomerang/Maxima is somewhat subjective.
The bottom panels show the corresponding 
spectrum of the primordial curvature perturbation, normalized to
1 at the COBE scale, against
 the  scale $\ell(k)$ probed by the cmb anisotropy.
The shortest scale shown (biggest $\ell$) corresponds to $k^{-1}\simeq
10 h^{-1}\Mpc$, at which the galaxy  data $\widetilde \sigma_8$ and
$\widetilde \Gamma$ apply.}
\label{fig2}
\end{figure*}
\end{center}

\begin{table*}
\begin{center}
\begin{minipage}{140mm}
\caption{Fit  of the $\Lambda$CDM model to presently available data,
assuming reionization when a fraction $f=10^{-2.2}$ of matter
has collapsed. (Corresponding redshift at best fit is $z\sub R=18$).
The scale-independent 
spectral index $n$ is a parameter of the model, and so are
the next three quantities. Every quantity except $n$ is 
a data point, with the value and uncertainty listed in
the first two rows. The result of the  least-squares fit is given in the
lines three to five.  All uncertainties are at the nominal 1-$\sigma$
level. The total $\chi^2$ is 6.3  with  three  degrees of freedom.}
\begin{tabular}{|c|c|ccc|cccc|}
\hline
& $n$ & $\omb h^2$ & $\Omega_0$ & $h$ 
&$\widetilde \Gamma$ & $\widetilde \sigma_8$ & $\sqrt{\widetilde C_\ell\su{1st}}$ &
$\widetilde C_\ell\su{2nd}/\widetilde C_\ell\su{1st}$\\
data & --- & 0.019 & 0.35 & 0.65 &
 0.23 & 0.56 & $74.0\muK$ & 0.38 \\
error & --- & 0.002 & 0.075 & 0.075  & 0.035 & 0.059 & $5\muK$
& $0.06$
 \\
fit & $0.987$ & $0.021$ & $0.38$ & $0.62$
 & 0.19  & 0.56  & $70.8\muK$ & 0.49\\
error & 0.051 & 0.002 & 0.06 & 0.05  & --- & --- & --- & --- \\
$\chi^2$ & --- & 0.9  & 0.2 & 0.2 &
 $1.3$ & $0.002$ & $0.4$ & 3.3 \\
\hline
\end{tabular}
\end{minipage}
\end{center}
\end{table*}

The fit minimizes $\chi^2$
with the assigned error  bars.
The   data set is the one given in the first two rows of Table 1,
plus the accurate value 
 provided by COBE at the relevant  scale $k\sub{COBE}$,\footnote
{This  value has a slight $\Omega_0$ dependence which we include. The
small uncertainty is ignored,  because  including the COBE normalization
in the fit returns values practically indistinguishable from the 
central one.}
\be
\frac25\calp_\calr^{\frac12}(k\sub{COBE}) = 1.94\times 10^{-5}
\label{cobenorm}
\,.
\ee
This 
 data set is the same as for the earlier fit, except that the height of the 
first peak is now taken from the Boomerang/Maxima data, and the
ratio of second to first peak height from the same source is now included.
For both peaks, we used the pair of data points nearest to the expected
peak position, one point from each of the two data sets. The random and
systematic errors for each point  were added in quadrature, and then
the weighted average was taken. 
The theoretical  peak heights
were in both cases taken from the output of the CMBfast package
\cite{CMBfast},
linearly interpolated as described earlier \cite{cl99} for the first peak.

Ours is the first fit which  takes account of 
both Boomerang and Maxima data,
 and which at 
the same time is global  in 
that there is an attempt to include in some form all data which is likely
to significantly constrain the model.\footnote
{After the original version of this paper appeared on the archive,
a more-or-less global fit did appear \cite{teg00}, which will
be discussed later.}
 Apart from cmb data,   we include 
the summaries of data on the galaxy correlation function and the
 cluster abundance 
 provided by the quantities
$\widetilde \Gamma$ and $\widetilde \sigma_8$,
admittedly subjective
estimates 
 of
$\Omega_0$ and $h$ (based on observations
\cite {likely1,likely2,likely3}
 that have nothing to do with the large-scale structure),
 and the `low-deuterium'
nucleosynthesis estimate of the baryon density \cite{osw,subir}.\footnote
{The alternative `high deuterium' estimate corresponds to a much lower
baryon density, which is disfavoured by the Boomerang/Maxima data.}
 Of course, we recognize that our choice of data points and error bars
is subjective.  For one thing, much of the uncertainty is systematic
making the minimization  of $\chi^2$ not strictly justified. 
For another, the device of representing many different measurements
by a single error bar loses information. In particular, we have dropped
measurements of the 
cmb anisotropy away from the first and second peaks, which are included
for instance by \cite{teg00}. 
Nevertheless,
 it seems to us reasonable to prefer some kind
of  global fit over fits  that arbitrarily keeps only  selected pieces 
of information such as, for instance, the cmb anisotropy. 
  Moreover, as can be seen from Fig. 2, we can  to some extent 
justify our procedure {\it a posteriori} observing that our best fit 
results are giving a good interpolation also of the cmb data we are 
neglecting (in the second peak region our fit is out by two standard 
deviations, but the situation does not become worse as one moves away 
from the peak):
this is surely not a chance, but due to the fact that 
the $\Lambda$CDM model gives a good account of the shape
of the peaks, so that the most important constraint comes indeed from the
peak heights.

\begin{center}
\begin{figure*}
\leavevmode
\epsfysize=7.5cm \epsfbox{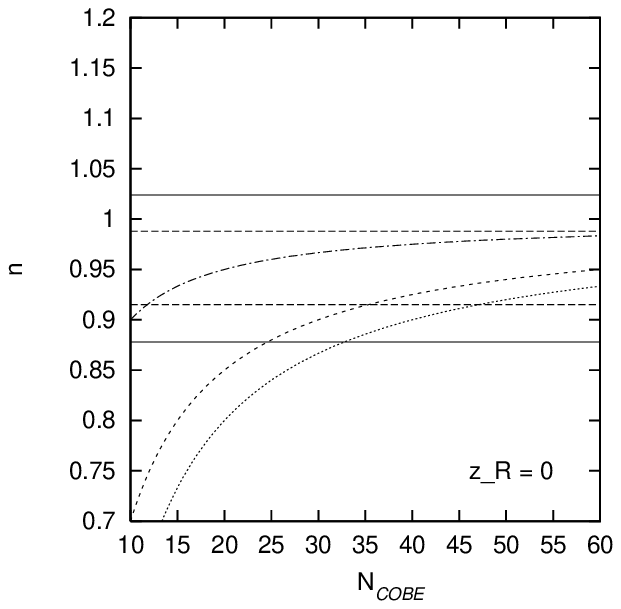}
\epsfysize=7.5cm \epsfbox{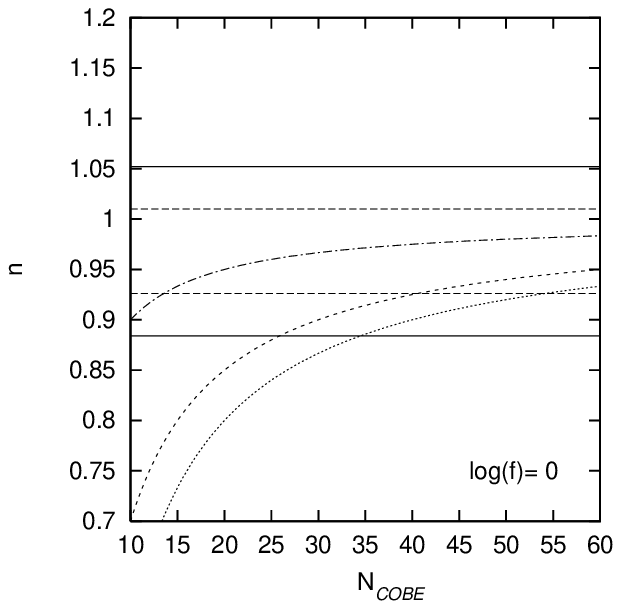}\\
\leavevmode
\epsfysize=7.5cm \epsfbox{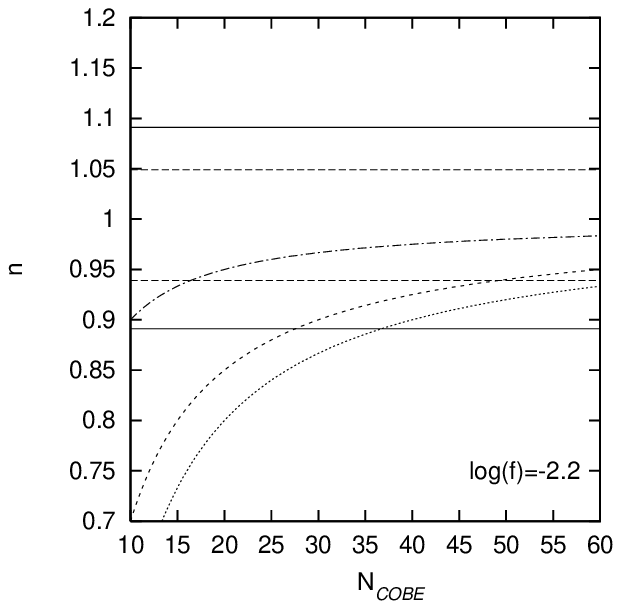}
\epsfysize=7.5cm \epsfbox{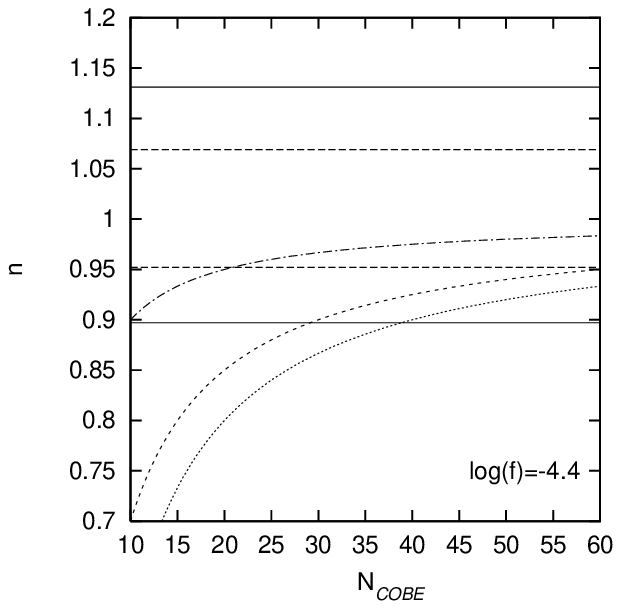}\\  
\caption{The horizontal lines show the 
 1- and 2-$\sigma$ bounds on $n$, with different panels corresponding to
different assumptions about the epoch of reionization. Also shown is the 
 dependence of $n$ on $N\sub{COBE}$, according to some of the models shown
in Table 2.
From top to bottom these are the logarithmic potential, new inflation
with $p=4$, and new inflation with $p=3$.
Significant lower bounds on $N\sub{COBE}$ are obtained for the
new inflation models. Taken  seriously, the 
   1-$\sigma$ bound   would practically rule out the $p=3$ model.}
\label{fig3}
\end{figure*}
\end{center}

Before describing our results, we want to 
describe our treatment of the reionization redshift $z\sub R$. 
Previously  reported
fits regard   $z\sub R$  as a parameter, to be either fixed at 
some reasonable value, or else to be included in the fit.
We prefer  an estimate of $z\sub R$ provided by the
$\Lambda$CDM model itself. Apart from having the virtue of 
keeping information which otherwise is lost, the use of this estimate
 will lead to a more realistic
lower bound on $n$. The estimate is obtained by 
 assuming  that reionization occurs
when some fraction $f\ll 1$ of the matter has collapsed into 
gravitationally bound structures, that epoch being estimated from
 the Press-Schechter approximation. We have considered values
of $f$ in the reasonable range \cite{llreion}
 $-4.4 <\log_{10} f <0$, with the result shown in the right hand panel
of Figure \ref{fig1}.\footnote
{The result for $10^{-1} <  f< 1$ is not shown, because in the approximation 
that we are using  \cite{cl99} it is  the same
as for the case $f=10^{-1}$.} When $f$ decreases  over this range, the
value of $z\sub R$ corresponding to the best fit increases from
$10$ to $26$. For comparison, we show in the left-hand panel of Figure
\ref{fig1} the result with $z\sub R$ fixed at various values.
Over the range $10<z\sub R< 26$, the upper bound on $n$ is similar
to the one which fixes $f$ (at the value reproducing $z\sub R$ at
best fit). In contrast, the lower bound on $n$ depends strongly on 
$z\sub R$, but relatively  weakly on $f$. The reason is that with fixed $f$
in our adopted range, low values of $n$ give low values of $z\sub R$.
 At the same time, the dependence of our result on the 
assumption concerning  reionization is not entirely insignificant.
Representative cases are given in the panels of Figure \ref{fig3},
along with some theoretical predictions that we shall discuss later.

 Taking
the central value $f=10^{-2.2}$, 
the  best-fit parameters are shown in Table 1.
The calculated data points are all  within one standard deviation
 or so of their 
observed values, except for the height of the second peak relative to
that of the first which is high by almost two  standard deviations, and
the  $\chi^2$ per degree of freedom is still perfectly reasonable.
The calculated cmb anisotropy is compared with the full Boomerang/Maxima 
data set in Figure \ref{fig2}, where the spectrum
of the primordial curvature perturbation is also shown.

We have also investigated the effect of omitting part of our data set.
First, we omitted the
data point for $\Omega\sub B h^2$ (coming from
 nucleosynthesis), 
and found  a best fit value $\Omega\sub B h^2\simeq 0.29$ in qualitative
agreement with other analyses \cite{boommax07333,teg00}. 
This is five standard deviations
higher than our the  data point, which in 
our view means that the fit is of little interest.
Second,  we omitted $h$ and/or
$\Omega_0$. Omitting just one of them makes little difference, because
their best-fit values are strongly correlated, but omitting both of them
again leads to values far away from the expectation
($\Omega_0\simeq 0.7$ and $h\simeq
0.45$) so that this  fit too is
 of little interest. 

Finally,  we investigated the effective of omitting
one or both of the large-scale structure data points
$\widetilde \Gamma$ and
$\widetilde \sigma_8$. Eliminating both leads to 
 best-fit values for $\widetilde \Gamma$ and
$\widetilde \sigma_8$ which are respectively three and four standard
deviations below the data.
 Here again, we take the view that such a fit is of little interest. 
(For the record, the fit gives $n=0.89\pm 0.07$, in 
qualitative agreement with another analysis \cite{mkr}.) 
Omitting just one of them
makes little difference to the fit, because they are again strongly
correlated. In particular, lowering $n$ lowers both the magnitude
of the spectrum on the relevant scales (measured by $\widetilde
\sigma_8$) and its slope (measured by $\widetilde \Gamma$).

Our fit is  similar to another one  \cite{teg00} (to be referred to as
TZH), with two
important  differences. First, the TZH data set 
 did not include  $\widetilde\sigma_8$ (nor any other constraint on the
normalization of the spectrum in the regime of large scale structure),
while it replaced our $\widetilde \Gamma$ by a 
fit to the shape of the  galaxy correlation function from a recent
infrared survey \cite{pscz}.\footnote
{ As mentioned earlier, 
TZH  also use a large data set of cmb anisotropy measurements,
but this difference from our treatment seems  to be less  crucial. }
The other difference is  that the reionization redshift was left as a 
free parameter, which in accordance with our finding makes the best fit
value practically zero. With this data set, the
 best fit of TZH gives parameters similar to ours, with spectral index
$n=0.91\pm0.05$ (1-$\sigma$) to be compared with our $z\sub R=0$ result
$n=0.95\pm0.03$. Using the best-fit parameters of TZH,  we find
$\widetilde \Gamma = 0.17$ and $\widetilde \sigma_8= 0.48$, both 
significantly lower than the data points we assigned to these quantities.
We believe that this is  the main reason for 
 the lower value of $n$ found by TZH.\footnote
{Ignoring the differences
in the other parameters, changing $n$ from the TZH value to our value
would  raise the spectrum at $k^{-1}=8h^{-1}\Mpc$ by $9\%$.
Roughly speaking, this  raises 
$\widetilde \sigma_8$ by the same amount, taking us from the TZH value
 $0.48$ to a value $0.53$ which is in reasonable agreement with the result
of our fit.}
It is interesting that the PSCz survey used by TZH is well-fitted by
a relatively low value of $\widetilde \Gamma$, which  corresponds
more or less to the one produced by our global fit. This may indicate
that the slope of the galaxy correlation function used by TZH is more
accurate than the older estimate that we (and many other authors) used.
Still, had  TZH  used $\widetilde\sigma_8$ as a data point,
their best fit to the slope of  the galaxy correlation function
 would have been somewhat higher, because of the correlation between the fitted
value of that slope with the value of  $\widetilde \sigma_8$.
As a result their best-fit value of $n$ would have been higher.
Probably more significantly, it would also have been higher had
they used our procedure of making a 
realistic estimate of the reionization redshift, instead of allowing it
to float. We shall see that our higher value of $n$ becomes interesting,
in the context of certain  models of inflation suggested by effective
field theory.


\section{Models of inflation}

\begin{center}
\begin{table*}
\begin{minipage}{140mm}
\caption[predictions]{
Predictions for the spectral index $n$, in terms of the 
number of $e$-folds $N$ to the end of slow-roll inflation.
Ignoring the slight scale-dependence, $N=N\sub{COBE}$
is around $50$ with the standard cosmology.
 For rows two and three there is a maximum
amount of inflation, corresponding to $N\sub{max}$.
All constants are  positive,  and $p$ is an integer except for the
mutated model. (The  mutated model can also give a potential of the
same form as `new' inflation with any $p$  bigger than
$2$.)
In the  top and bottom rows  $n$ can be far  from 1, 
while   in the other cases it 
is  typically  close to 1.
}
\begin{tabular}{|lll|}
\hline
 & $V(\phi)/V_0$ & $\frac12 (n-1)$ \\
Positive mass-squared  & $1+\frac12 \frac{m^2}{V_0} \phi^2$ 
& $ \mpl^2 m^2/V_0$ 
\\
Self interaction ($p\geq 3$)
 & $1+c\phi^p$ &
$\frac{p-1}{p-2}\frac1{N\sub{max}-N}$ \\
Dynamical symmetry breaking   ($p\geq 1$)
 & $1+c\phi^{-p}$ &
$\frac{p+1}{p+2}\frac1{N\sub{max}-N}$  \\
Loop correction  &
$1+c \ln\frac\phi Q$ & $-\frac1{2N} $  \\
Mutated ($-1<p<\infty$)   & $1-c\phi^{-p}$ &
$-\(\frac{p+1}{p+2} \) \frac1 N$   \\
`New'  ($p\geq 3$)  & $1-c\phi^p$ &
$-\(\frac{p-1}{p-2} \) \frac1 N$   \\
Negative mass-squared & $1-\frac12 \frac{m^2}{V_0} \phi^2$ 
& $ -\mpl^2 m^2/V_0$ \\
\hline
\end{tabular}
\end{minipage}
\label{table2}
\end{table*}
\end{center}

\subsection{The general framework}

Our   next objective is to compare the observational constraint
on $n$ with some  models of inflation. Before getting to the details,
it will be as well to describe the general framework within which we 
operate, since it radically differs both from the `reconstruction' program
 \cite{recon} and from a proposed  `small/large/hybrid' classification
of potentials \cite{dkk,mkr}.

Inflation is supposed to do two separate jobs.
One 
is to evolve  the Universe 
 from a generic initial condition at, presumably,
the Planck scale, without either collapsing or becoming empty.
The other  is to set, after inflation is over,
 the  initial conditions that describe the observable  Universe,
in particular the primordial curvature perturbation.
The second job is done  during the
last fifty or so $e$-folds of inflation, while observable scales
are leaving the horizon. 
To produce the nearly scale-invariant perturbation that we see,
 inflation 
during this era presumably has to be of the slow-roll variety, which 
implies that $\rho^{1/4}$ is at least a couple of orders of magnitude
below the Planck scale (\eq{cobev} below). 
For our  purpose, a `model of inflation' is a model of this era,
which alone is accessable to observation.

At the most primitive level,  a model of inflation is a form for the 
potential during inflation, and a specification of the field
value at the end of slow-roll inflation. 
(In hybrid models, the field value at the end of slow-roll inflation
is not determined by the form of the potential during inflation,
because the end corresponds to the
 de-stablisation of some non-inflaton field.)
At this primitive 
level, though, one has complete freedom in choosing the form
of the potential during inflation,
and  consequently very little predictive power.\footnote
{{}Even the relatively restrictive slow-roll paradigm presented below
leads  only to the gravitational wave  constraint
$r=-6.2 n\sub T$ and the flatness conditions 
 $|n-1|\ll 8$, $|n\sub T|\ll 2$.}
In order to reduce the freedom, one therefore  looks
for guidance to effective field theory.

Effective field   theory provides
the framework for the Standard Model 
of particle physics and for  its phenomenological  extensions.
It is supposed to be valid on energy scales far below the 
 ultra-violet cutoff $\Lambda\sub{UV}$, which is at most of order
 the Planck scale. In this regime, the  unknown physics beyond the cutoff
is   ignored (in a renormalizable theory)
 or else encoded by the inclusion of the leading  non-renormalizable
term(s). In contrast, an attempt to use the effective theory on scales
approaching the cutoff would require an infinite number of non-renormalizable
terms, leading to a complete loss of predictive power.
In the present context, the relevant non-renormalizable terms are contributions
to $V$ of the form
$\lambda_n \phi^{4+n}/\Lambda\sub{UV}^n$, with expected coefficients
 $\lambda_n\sim 1$.
In models based on effective field theory 
 the non-renormalizable terms are essentially
ignored, and taking optimistically 
$\Lambda\sub{UV}\sim \mpl$ this neglect requires 
 $|\phi|\ll \mpl$.\footnote
{In the usual case, that the
real inflaton field $\phi$ is the radial part of some complex
field,  the  origin $\phi=0$ can be  taken
as  the fixed point of the symmetries of the renormalizable field
theory. If the inflaton field is the angular part (a pseudo-Goldstone boson)
it runs over only a finite range, and 
its origin within this range is arbitrary.
Irrespective of the definition of the origin,
the neglect of non-renormalizable terms requires the {\em range}
of  $\phi$ spanned by relevant field values to be much less than
$\Lambda\sub{UV}$, and most of the following discussion goes through
if this requirement replaces the assumption the $\phi$ itself is small.
Note that the assumption of small $\phi$ is a minimal one, necessary 
to justify the  neglect of non-renormalizable terms but by no means
always sufficient \cite{treview}.}
(Sometimes one forbids non-renormalizable terms in 
$V(\phi)$ by invoking a suitable  symmetry, but according to current ideas
 this is likely to work only for a limited number of terms
\cite{treview,ewanglobal}.)

Instead of effective field theory, one may hope to use
 a  deeper theory like string theory, which would determine the coefficients
of all of the non-renormalizable terms. Such a theory might predict that
the coefficients of these  terms are very small (in Planck units), making it
easy to have inflation with very large field values.
This  approach does
yield some proposals for the potential of moduli 
\cite{treview,bd}, perhaps leading to inflation with a potential of the
form $V\simeq V_0-\frac12m^2\phi^2+\cdots$ (last row of Table \ref{table2})
which would require a field variation of order $\mpl$.
(A different stringy proposal 
 \cite{gia} does not lead to a viable model.)
This exception apart, it seems that at the present time predictive models
have to be based on effective field theory,
 in which the relevant values
of the inflaton field are  small on the Planck scale. 
 We do not know whether Nature has chosen this option, or has
instead chosen inflation with large field values,
 but we take the view that in the latter case there is at present no
theoretical guidance as to the form of the potential. Hence we focus on
the effective field theory models.

\subsection{Slow-roll inflation}
\label{slowroll}

Slow-roll inflation, with a single-component inflaton field,
 is described by 
the following basic set of formulas  \cite{treview,book}.\footnote
{Multi-component models
 require a different treatment, as do  models in which slow-roll
is briefly interrupted, and models with a quartic kinetic term
\cite{kinetic}.
The inflaton field is supposed to be canonically normalized, which
can always be achieved if the kinetic term has the usual quadratic
form.
Einstein gravity during inflation is assumed,
 which can usually be achieved by a conformal transformation even
if the underlying theory is non-Einstein.}
In these formulas, 
 $\phi$ is the inflaton field, and $V$ is its potential during inflation.
The other quantities are  the 
Planck masss
 $\mpl=2.4\times 10^{18}\GeV$,
 the scale factor of the Universe $a$, 
 the Hubble parameter  $H=\dot a/a$, and
  the wavenumber $k/a$ of the cosmological perturbations.
 We assume the usual flatness  conditions
\bea
\epsilon&\ll &1,\hspace{2em} |\eta|\ll 1 \label{flatness} \\
\epsilon&\equiv& \frac12\mpl^2(V'/V)^2 \nonumber\\
\eta&\equiv&  \mpl^2V''/V \nonumber
\,,
\eea
 leading to the slow-roll expression
$3H\dot\phi\simeq -V'$.

 The fundamental formula  giving  the spectrum of the 
curvature perturbation is
\be
\frac4{25}\calpr(k)
 = \frac1{75\pi^2\mpl^6}\frac{V^3}{V'^2} \,,
\label{delh}
\ee
where the 
 potential and its derivatives are
 evaluated at the epoch of horizon exit
$k=aH$. On the scale $k\sub{COBE}$,
the  COBE normalization \eq{cobenorm} requires
\be
V^{1/4} = .027 \epsilon^{1/4} \mpl
\,.
\label{cobev}
\ee

To work out the value of $\phi$ at the  epoch of horizon exit,  one uses
the relation
\be
\ln(k\sub{end}/k)\equiv N(k)
=\mpl^{-2}\int^\phi_{\phi\sub{end}} (V/V') \diff\phi
\,,
\label{Nofv}
\ee
 where $N(k)$ is actually the number
of $e$-folds from horizon exit  to the  end of slow-roll inflation.
The biggest scale of interest may be taken as 
 $k^{-1}\sub{COBE}$, which using the 
definition in
 our earlier work \cite{cl99}
corresponds to  $k^{-1}\sub{COBE} \simeq 730h^{-1}\Mpc$.
 Observations of the smaller scale cmb anisotropy
and galaxy surveys take us down to say $k^{-1}\sim 10 h^{-1}\Mpc$, 
corresponding to a change $\Delta N\simeq 4 $ to $5$.
At a given scale, $N$ depends on the post-inflationary evolution of the
scale factor, and using for definiteness the COBE scale it is usefully
written as
\be
N\sub{COBE} \simeq 60 - \ln\(\frac{10^{16}\GeV}{V^{1/4}}\) - 
\frac13\ln\(\frac{V^{1/4}}{T\sub{reh}}\)-N_0
\, .
\label{Ncobe}
\ee
In this expression, $T\sub{reh}$ is the  reheat temperature
after inflation, and $V$ is the potential at the end of inflation.
The  final contribution  $-N_0$ (negative in all reasonable cosmologies)
 encodes our ignorance about
what happens between this reheat  and nucleosynthesis.
In the conventional cosmology, with the relatively high inflation
scale occurring in most of our models, one  expects $N\sub{COBE}\sim 50$
 to $60$, but 
  late entropy release from  thermal inflation 
\cite{thermal} (or a low value of $V$) can  make  $N\sub{COBE}$ much lower.
In some of the inflation models that we shall consider, 
 our  bound on $n$ will lead to  a useful
lower bound  on $N\sub{COBE}$.

The spectral index $n$ is defined by
\be
n(k)-1\equiv \frac {\diff \ln \calpr}{ \diff \ln k}
\,,
\ee
{}and is given by 
 \eqs{delh}{Nofv} as \cite{ll92}
\bea
n-1 &=&  2\eta-6\epsilon
\,.
\label{nofv}
\eea
It defines the scale-dependence of the $\calp_\calr(k)$ leaving its
value at (say) the COBE scale as the only other quantity required to
completely specify it.

Finally, the spectrum of the primordial gravitational waves is characterized
by its contribution $r$ to the spectrum of the cmb anisotropy on the 
COBE scale (defined in
a certain approximation and measured in units of the contribution of the
curvature perturbation) and its spectral index $n\sub T$, which are
given by \cite{ll92}
\bea
r&=&12.4\epsilon \label{r}\\
n\sub T &=& - 2\epsilon \label{nt}
\,.
\eea

We are going to apply these slow-roll  equations to
models of inflation in which the relevant values of the inflaton field
are small. Quite generally, such  models 
 predict that $r$ is too small to observe in the foreseeable future,
 in accordance with the
assumption of our fit \cite{mygwave}. Indeed,
applying  \eq{Nofv}  to 
the range of  scales $\Delta N\simeq 4$ over which gravitational waves
affect  the cmb anisotropy, one finds
\be
\frac{r}{0.1} \sim \( \frac{\Delta \phi}{0.5 \mpl} \)^2
\,,
\ee
where $\Delta\phi$ is the corresponding change in $\phi$, and
$r\simeq 0.1$ is about the smallest signal that can be detected
by the PLANCK satellite. A detectable signal therefore requires
that the change in $\phi$ over relevant field values be large,
and therefore that the value of $\phi$ itself be large for at least
some relevant field values. We emphasize again that we have no idea
whether Nature has chosen small or large field values, and that we
focus on the small-field case because in our view only that case is
at present understandable from the point of view of effective field theory.

For future reference, we note that the converse of the above result
does not apply. The total change in $\phi$ after the COBE scale leaves
the horizon can be large, in a model where the
  change over the relevant four $e$-folds is small.
 As a result, large-field models do not necessarily
lead to significant gravitational waves. For instance, if
  $V$ depends linearly on  $\phi$, then
  $N(\phi)\propto   \phi$, and the change in $\phi$
after the COBE scale leaves the horizon is related to the 
change $\Delta \phi$ during four $e$-folds by
$\phi\sub{COBE}/\Delta\phi\simeq  N\sub{COBE}/4 \gg 1$.

\subsection{Some simple models}

Effective field theory, with non-renormalizable terms
essentially ignored, allows  only a few different types of term
for the variation of $V$. With the reasonable  assumption that 
one such term dominates over the relevant range of $\phi$,
we arrive at essentially the models displayed in Table 2.
Details of these models, with extensive  references and possible
complications, are given
in \cite{treview}. One of these complications is the 
possibility, considered by several authors,
 that two terms need to be kept over the relevant range of $\phi$.
While this can happen, 
the dominance of one term is the
generic situation in the sense that it holds
over most of the potential's parameter space.\footnote
{ Consider, for  instance, the case that 
there are just two parameters, corresponding to the overall normalization
of the two terms. At a given field value, parameter space  then consists
of a region where one term dominates and a region where the other term 
dominates, these regions being divided by a line corresponding to 
$50\%$ of each term. If we consider the cosmological range of field values,
and interpret the `dominance' of one term as say a factor of ten 
between them, the line becomes a band, but still a set of measure zero
compared with all of parameter space. Finally,  the cobe normalization
corresponds to a line in parameter space, which will generically cross the
band; only very exceptionally will the line lie within the band. 
  See for an example Fig. 1 of \cite{bcd}, which shows for a
specific model that the full potential is well approximated by a 
single term in the region of dominance.   }
 
 When the COBE normalization is imposed on the
prediction, the requirement that the  relevant  values of the 
inflaton field be small 
can generally be  satisfied  with physically reasonable values  of the
parameters. An exception is the potential $V=V_0-\frac12m^2\phi^2
+\cdots$, which requires $\phi\sim\mpl$ at the end of inflation
(see below). 
The only other significant exception is the
logarithmic potential, which requires  $\phi\sim\mpl$ when cosmological
scales leave the horizon,  if the coupling is unsuppressed as in
the case of $D$-term inflation.

Given the restriction on $\phi$, the flatness conditions (\ref{flatness})
require that $V_0$ dominates the potential in all of the models,
 leading to simple expressions for $\epsilon$ and $\eta$.
The  contribution of
 gravitational waves is negligibly small in all of them,
 and 
 the formula for  $n$ is well approximated by
\be
n-1=2\eta
\label{nofvapprox}
\,.
\ee
The resulting prediction for $n$ is 
 shown in Table 2. Except in the first and last rows, the prediction
  depends on $N$ and is therefore scale-dependent. However, since
 $n$ is constrained to be close to 1,
the   scale-dependence is negligible
 over the cosmological
range $\Delta N\sim 4$ \cite{cl99}, and accordingly one may set
$N=N\sub{COBE}$.
The bottom two  rows correspond to single-field 
 inflation with a mass term or a 
self-interaction dominating, and an unspecified term stabilizing the potential
after inflation. The other rows correspond to hybrid 
inflation, where a non-inflaton field is responsible for most of the 
potential during inflation. 
The loop correction is the one which arises
with spontaneously broken global supersymmetry, as for example in 
`$D$-term inflation'. The ranges of $p$ are the ones in which the prediction
for the spectral index holds, and  they can be achieved in effective
field theory with at most a single 
non-renormalizable term.

The strongest  prediction comes from the 
models  giving $n-1\propto 1/N$. It is 
 shown in  Figure \ref{fig3} for  the three most popular versions of these 
models, along with the observational
 bounds on $n$.
In the `new' inflation models 
there is a non-trivial lower bound on $N$, which would almost
exclude the $p=3$ model if the 1-$\sigma$ bound were taken seriously.

Another case of interest is the potential $V=V_0 -\frac12 m^2\phi^2
+\cdots$. 
More or less independently of the additional terms which stabilize
the potential, the vev of $\phi$ is  $\langle\phi\rangle\sim\sqrt (2V_0/m^2)
=[2/(1-n)]^{1/2} \mpl$.
Depending on the nature of $\phi$, this kind of inflation has been termed 
`natural',      `topological' and `modular' (see for instance
\cite{bd} for a recent espousal of modular inflation).
In all cases 
the model is regarded as  implausible if  $\langle\phi\rangle$ is much
bigger than $\mpl$, which means that it is viable only if 
$n$ is not too close to 1.
 Our
 2-$\sigma$ bound
$n\gsim 0.9$ implies $\langle\phi\rangle \gsim 4.5\mpl$, which 
may perhaps be regarded as already disfavouring these models.

\subsection{Alternative views}

The view we have taken is 
different from the one espoused
in \cite{dkk,mkr}. These authors consider
 the potentials in first, second, third and last rows of  Table 2
(and a linear potential)  but unlike us they
take such  field-theoretic forms of the potential
  seriously even at $\phi\gsim\mpl$.
In particular, they consider the limit where the constant term 
is negligible, corresponding to monomial potentials like
$V\propto \phi^2$ \cite{chaotic}.\footnote
{With these particular examples in mind, they divide the $n$-$r$ plane
into regimes corresponding to $\eta<0$, $0<\eta<2\epsilon$ and $2\epsilon
<\eta$, labeling them respectively the (non-hybrid)
small-field, (non-hybrid) large-field and hybrid regimes. 
These designations do not have general validity,
and seem indeed to be confined to the particular examples
already mentioned.}
This procedure
 allows a significant
gravitational wave contribution  $r$, and a wide range of $n$ for each $r$.
Therefore, to delineate the allowed region of parameter space,
the authors of \cite{mkr}  allow both $n$ and $r$ to vary.
We, in contrast,  consider only small field values,
leading to negligible $r$ which we set equal to zero.
We  take  the view that in
 the regime 
$\phi\gsim \mpl$,  a single-power  form for the variation
of $\phi$ is  no more likely 
 than any other, and therefore that
the potentials in Table 2 have no special status 
in this regime.

 While  the assumption of a  single power  seems too restrictive,
it  might reasonably be argued  that a 
  combination of two or three powers
(say positive integral powers)
  will be sufficiently flexible to cover a useful range of potentials.
Why, then, in the large-field regime,
do  we not wish to  focus in  particular on the case that one 
power  dominates, as we did for the small-field regime?
Our  answer illustrates beautifully the difference of view that we take
in these  two regimes. In the small-field regime, each power listed
in Table \ref{table2} has a more or less definite  physical origin;
for instance,  positive powers correspond to different self-coupling terms
for the inflaton. In that situation, it 
 seems  reasonable to regard the coefficients of these powers 
as parameters which,
 in the present state of theory, have more or less equal  prior
probability. Then, over most of parameter space (and even over most of the
restricted region allowed by the COBE normalization) just one power
dominates, making this the most likely situation.
In the large-field case, on the other hand,
we argue that a single power has  no special significance, and neither do
the parameters of an expansion consisting of several powers.
Instead of single powers,  one could choose a new  basis
consisting of  linear combination of powers (say, Legendre polynomials instead
of monomials)  leading to new `parameters'
which would have to be chosen specially to reproduce a single power.
This leads us back to the position stated earlier, that in our view
there is no theoretical guidance as to the form of the potential
in the large-field regime.

Precisely this  view  provided the starting
point of the `reconstruction' program reviewed in \cite{recon}.
The idea here is that, if significant gravitational waves are observed,
then \eqsss{delh}{nofv}{r}{nt},
 and more accurate versions \cite{sl93} involving
one higher derivative
of  $V$, may  allow $V$ to be reconstructed over the very limited
 range of $\phi$
corresponding to horizon exit for cosmological scales. 
We have no comment on this purely phenomenological 
approach, except to emphasize  that it works only if $r$ is big enough
to measure.

\section{Running mass models}

\begin{figure*}
\centering
\leavevmode
\epsfysize=7.5cm \epsfbox{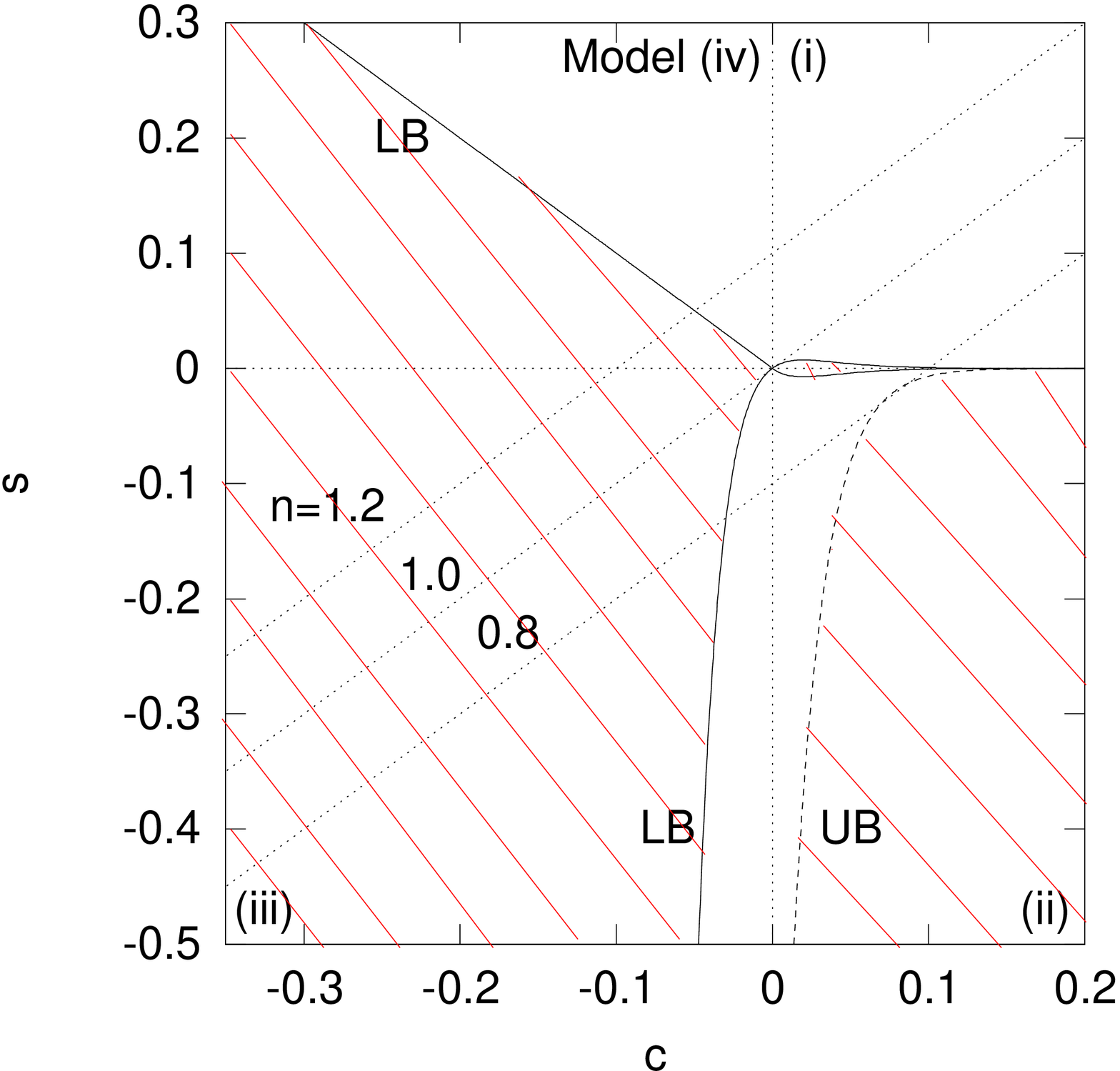}
\epsfysize=7.5cm \epsfbox{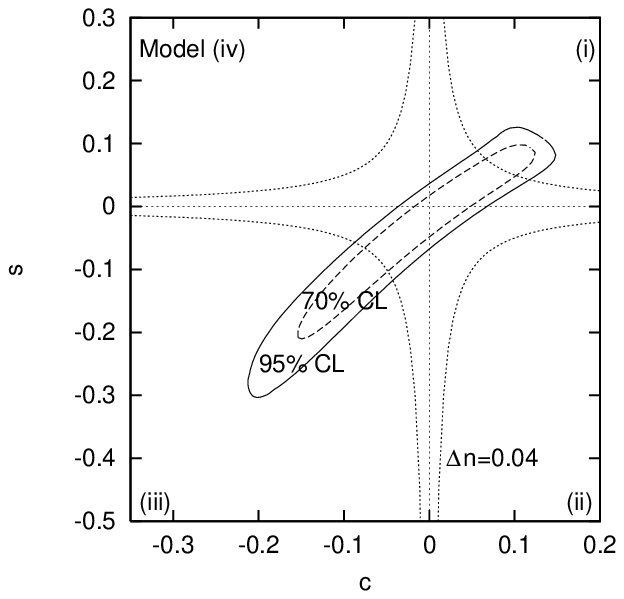}
\caption[sc-fig2]{
The  parameter space for the running mass model. As discussed in \cite{cl99}
the model comes in four versions, corresponding to the four quadrants
of the parameter space.
In the left-hand panel,  the straight lines 
corresponding to $n\sub{COBE}=1.2, 1.0$ and $0.8$, and the
 shaded region is disfavoured on  theoretical grounds.
In the right-hand panel, we show the region allowed by observation,
in  the case that reionization occurs when  $f\sim 1$.
To show the scale-dependence of the prediction for $n$, we also show in 
this panel the branches
of the hyperbola $8sc=\Delta n\equiv n_8-n\sub{COBE}$, for the reference
value $\Delta n=0.04$.}
\label{s-c-f1}
\end{figure*}

\begin{table*}
\begin{center}
\begin{minipage}{140mm}
\caption{Fit  of the $\Lambda$CDM model to presently available data.
The  scale-dependent spectral index is given by  \eq{nofsc} with $c=0.1$.
Free parameters are $n\sub{COBE}=1+2(s-c)$, and the next three 
quantities in the Table. Reionization is taken to occur 
 when a fraction $f=10^{-2.2}$ of matter
has collapsed. (The corresponding redshift at best fit is $z\sub R=21$.)
 Every quantity except $n\sub{COBE}$ is 
a data point, with the value and uncertainty listed in
the first two rows. The result of the  least-squares fit is given in the
lines three to five.  All uncertainties are at the nominal 1-$\sigma$ level. 
The total $\chi^2$ is 8.4 with three degrees of freedom.}
\begin{tabular}{|c|c|ccc|cccc|}
\hline
& $n\sub{COBE}$ & $\omb h^2$ & $\Omega_0$ & $h$ 
&$\widetilde \Gamma$ & $\widetilde \sigma_8$ & $\widetilde C_\ell\su{1st}$ &
$\widetilde C_\ell\su{2nd}/\widetilde C_\ell\su{1st}$\\
data & --- & 0.019 & 0.35 & 0.65 &
 0.23 & 0.56 & $74.0\muK$ & 0.38 \\
error & --- & 0.002 & 0.075 & 0.075  & 0.035 & 0.059 & $5\muK$
& $.06$
 \\
fit & 0.94 & $0.021$ & $0.40$ & $0.59$
 & 0.19  & 0.53 & $67.6\muK$ & 0.49\\
error & 0.04  & 0.002 & 0.05 & 0.05  & --- & --- & --- & --- \\
$\chi^2$ & --- & 0.9  & 0.4 & 0.6 &
 $1.2$ & $0.2$ & $1.6$ & 3.5 \\
\hline
\end{tabular}
\end{minipage}
\label{table4}
\end{center}
\end{table*}

We also considered the case of running mass inflation
models,
 which give a spectral index 
with potentially strong scale
dependence. The potential in this case is
corresponds to a loop correction in the context of softly broken
global supersymmetry, and is 
 of the form
\be
V = V_0 \(1+ {1\over p} c\phi^p \ln(\phi/Q) \)
\,.
\ee
The case  $p=2$ corresponds to the  a renormalizable interaction,
which alone has been studied so far
\cite{st97,st97bis,clr98,cl98,c98,rs,cl99}.
It  gives a spectral index of the form
\bea
{n(k)-1\over 2} &=& s \exp (c \Delta N(k)) - c \label{nofsc}
\\
\Delta N (k) &\equiv& N(k\sub{COBE})- N(k)
\eea
These models invoke the loop correction coming from softly, as opposed
to spontaneously, broken renormalizable global supersymmetry. In the simplest
scenario, $c$ is essentially the coupling strength of the field in the loop,
which is expected to be of order $0.1$ to (say) $0.01$ in the case of 
a gauge coupling. 

Because of the possibly strong scale-dependence of the curvature
perturbation, our procedure of calculating the reionization
redshift in terms of the fraction $f$ of matter collapsed
becomes crucial. The results are insensitive to $f$ in the
reasonable range $f\gsim 10^{-4}$, which would not at all be the case
if we fixed instead the reionization redshift.

In Figure \ref{s-c-f1} we show the allowed region of parameter
space, with $f=1$. We see that $c\sim 0.1$ is indeed allowed,
and Table 4 we show the result of a fit with $c$ fixed
at this value, with a central value $f=10^{-2.2}$.
The corresponding cmb anisotropy
and curvature spectrum are shown in 
Figure \ref{fig2}. Comparing with the corresponding figures
for the case of a scale-independent spectral index,
the  scale-dependence generated by the coupling $c=0.1$
is clearly visible. Although   observation
 cannot yet distinguish clearly
 between the two cases, it  is likely to do so in the future,
deciding whether the inflaton has an unsuppressed coupling 
in this type of model. Note that, in contrast with the earlier fit
\cite{cl99}, the maximum allowed value of $n\sub{COBE}$ is now too small
for over-production of primordial black holes at the end
of inflation  \cite{lgl00} to be a problem.

\section{Conclusion}

Continuing earlier work \cite{cl99}, we
 have  fitted the $\Lambda$CDM model to a
global data set, assuming that a gaussian  primordial curvature perturbation
is the only one. The data set now includes heights of the first two
peaks in the cmb anisotropy,  derived
from the Boomerang and Maxima data.
We focus  on the spectral index $n$, specifying the shape
of the curvature perturbation, considering separately the case of 
a practically scale-independent spectral index, and the scale-dependent
spectral index predicted by running mass inflation models.
In contrast with other groups, we calculate the reionization epoch 
within the model on the assumption that it corresponds to the 
epoch when some fraction $f$ of the matter collapses, the results
being only mildly dependent on  $f$ in the reasonable range $f\gsim 10^{-4}$.

 For the scale-independent case, the
 bounds on $n$ are given in Figure \ref{fig3} for some typical
values of $f$, and for the case of no reionization.
In the same Figure, the bounds are 
 compared  with the prediction of some forms 
 of the inflationary potential which are suggested by
 effective field theory.
 The prediction depends on the number
of $e$-folds $N$ of inflation after cosmological scales leave the 
horizon, where $N\lsim 60$
 depends on the post-inflationary cosmology. For two of the models,
the constraint on $n$ rules out a significant portion of the
$n$-$N$ plane. 

In the case of running mass models, the scale-dependent spectral index
depends on parameters $s$ and $c$, the latter being related to the
inflaton coupling which produces the running. We have delineated
the allowed region in the $s$-$c$ plane. An unsuppressed coupling
$c\sim 0.1$ is allowed by the data, leading to a noticeable
scale-dependence of the spectral index. The  fit with $c=0.1$
 is less good than with a scale-independent spectral index,
but still acceptable.

\section{Acknowledgments}

We would like to thank A. D. Linde for useful comments on the first
version of this paper.

\newcommand\pl[3]{#3, Phys. Lett., #1, #2}
\newcommand\pr[3]{#3, Phys. Rep., #1, #2}
\newcommand\prl[3]{#3, Phys. Rev. Lett., #1, #2}
\newcommand\prd[3]{#3, Phys. Rev., D#1, #2}
\newcommand\ptpl[3]{#3, Prog. Theor. Phys., #1, #2}
\newcommand\rpp[3]{#3, Rep. on Prog. in Phys., #1, #2}
\newcommand\rmp[3]{#3, Rev. Mod. Phys., #1, #2}
\newcommand\jhep[2]{#2, JHEP, #1}
\newcommand\grg[3]{#3, Gen. Rel. Grav.,  #1, #2}
\newcommand\mnras[3]{#3, MNRAS, #1, #2}
\newcommand\apj[3]{#3, ApJ, #1, #2}
\newcommand\nature[3]{#3, Nature, #1, #2}
\newcommand\ps[3]{#3, Physical Scripta, #1, #2}
\newcommand\science[3]{#3, Science, #1, #2}

\end{document}